\documentclass[
	aps, prd, reprint, a4paper,
	amsmath, amssymb, amsfonts, eqsecnum,
	superscriptaddress,
	showpacs, showkeys,
	nofootinbib, %longbibliography,
]{revtex4-1}

\usepackage{ifthen}
\newboolean{prd}
\setboolean{prd}{false}
\newboolean{arxiv}
\setboolean{arxiv}{false}

\newboolean{notprd}
\setboolean{notprd}{true}
\ifprd
\setboolean{notprd}{false}
\fi
\newboolean{notarxiv}
\setboolean{notarxiv}{true}
\ifarxiv
\setboolean{notarxiv}{false}
\fi

\usepackage{xcolor}
\usepackage{dsfont}
\usepackage{graphicx}

\ifnotprd
\xdefinecolor{mylinkcolor}{rgb}{0,0,0.5}
\usepackage[
	bookmarksnumbered, bookmarksopen, bookmarksopenlevel=2,
\ifnotarxiv
	breaklinks=true,
\fi
	colorlinks=true, filecolor=mylinkcolor, citecolor=mylinkcolor,
	linkcolor=mylinkcolor, urlcolor=mylinkcolor, menucolor=mylinkcolor,
]{hyperref}
\else

\fi

\usepackage[normalem]{ulem}

\def\vct#1{\mathbf{#1}}

\def\dd{\mathrm{d}}

\allowdisplaybreaks

\begin{document}
% for PRD one has to get rid of hyperref
\ifnotprd
\hypersetup{
	pdftitle={Observables for the motion of a test-mass in a post-Newtonian approximated Kerr spacetime field to leading order
 quadratic in spin for an inclinated orbital plane},
	pdfauthor={Steven Hergt, Gerhard Schaefer, Abhay Shah}
}
\fi

%\title{Observables for the motion of a test-mass in a post-Newtonian approximated Kerr spacetime field to leading order quadratic in spin for an inclined orbital plane}
\title{Observables of a test-mass along an inclined orbit in a post-Newtonian approximated Kerr spacetime to leading-order-quadratic-in-spin}

\author{Steven Hergt}
\email{steven.hergt@uni-jena.de}
\affiliation{Theoretisch--Physikalisches Institut, \\
	Friedrich--Schiller--Universit\"at Jena, \\
	Max--Wien--Platz 1, 07743 Jena, Germany, EU}

\author{Abhay Shah}
\email{abhay@weizmann.ac.il}
 \affiliation{Benoziyo Center for Astrophysics, \\
	Weizmann Institute of Science, \\
	P.O. Box 26, Rehovot 76100 Israel}

\author{Gerhard Sch\"afer}
\email{gos@tpi.uni-jena.de}
\affiliation{Theoretisch--Physikalisches Institut, \\
	Friedrich--Schiller--Universit\"at Jena, \\
	Max--Wien--Platz 1, 07743 Jena, Germany, EU}

\date{\today}

\begin{abstract}
The orbital motion is derived for a non-spinning test-mass in the relativistic,
 gravitational field of a rotationally deformed body not restricted to the equatorial plane or spherical orbit. 
The gravitational field of the central body is represented by the Kerr metric, expanded to second post-Newtonian
 order including the linear and quadratic spin terms. The orbital period, the intrinsic periastron
advance, and the precession of the orbital plane are derived
with the aid of novel canonical variables and action-based methods.    
\end{abstract}


\maketitle
%\tableofcontents
%Here come some important references for the paper: First calculation of orbital elements linear in spin with %inclinated orbital plan \cite{Damour:Schafer:1988},
%different approach by Hackmann to calculate a differing post-Newtonian periastron shift by approximating the exact solutions
% \cite{Hackmann:2010,Hackmann:Lammerzahl:2012},
%for the Schwarzschild case alone see \cite{Scharf:2011}, the comparison of the orbital elements for the spin %orbital momentum aligned case is made with
% \cite{Tessmer:Hartung:Schafer:2012}, first introduction of carter constant \cite{Carter:1968}, for recent %handling of Carter constant see \cite{Flanagan:Hinderer:2007}
% and \cite{Will:2009}, the Kerr metric was introduced in \cite{Kerr:1963} and written in Boyer-Lindquist %coordinates in \cite{Boyer:Lindquist:1967}, for action approach
%with spin see %\cite{Steinhoff:2011,Steinhoff:Schafer:2009:2,Hergt:Steinhoff:Schafer:2011,Tessmer:Steinhoff:Schafer:2013}. For %orbital parametrization see
% \cite{Konigsdorffer:Gopakumar:2005}, for introducton of Hill variables see \cite{Schneider:Cui:2005}, for %gravitational self force handling see
% \cite{LeTiec:Barausse:Buonanno:2012}, for contour integration see \cite{Sommerfeld:1951}, for periastron shift %calculation also see \cite{Landau:Lifshitz:Vol2:2},
% for action-angle variables see \cite{Goldstein:1981}, for introduction of Mino
%time see \cite{Mino:2003}, for analytical solutions prior to Hackmann see \cite{Landau:Lifshitz:Vol1}.

\section{Introduction}

The motion of spinless particles in the gravitational field of a rotating body is an interesting and involved problem if the
motion is not restricted to the equatorial plane of the central body. 
Ever since the discovery of quasi-periodic oscillations (QPOs) and horizontal-branch oscillations in X-ray sources such 
as neutron stars, micro-quasars or black holes, theorists have tried to explain the presence of these oscillations by identifying
 their frequencies with those of nodal precession frequency of `blobs' in the accreting disks or Keplerian frequencies or periastron
 advance frequencies of the accreting rings. Since the mechanism that leads to these oscillations are associated with processes in 
the accretion disk, earlier calculations have been restricted to either close-to-equatorial plane or spherical or nearly circular
 orbits. Merloni in \cite{Merloni:1999} calculates the 
Lense-Thirring precession frequency of test-masses restricted to spherical orbits (constant $r$) and identifies them with 
the frequencies of QPOs from black hole sources. Sibgatullin in \cite{Sibgatullin:2001} calculates the nodal and 
periastron-rotation frequencies for nearly circular, non-equatorial orbits. Comparison of the observed and theoretical 
frequencies, once the relation between them confirmed, also act as a tool to test the general theory of relativity in 
strongly gravitating regions. Exact analytic solutions for arbitrarily oriented orbits are known in
 case of rotating black holes \cite{Schmidt:2002,Fujita:Hikida:2009,Hackmann2:2010,Hackmann:Lammerzahl:2012} but the Newtonian analog or the post-Newtonian (pN) expansion 
of the results, say periastron
 shift, is not straightforward, see \cite{Sibgatullin:2001}.
 Exact expression for the periastron advance of a test-mass in the gravitational field of a
 non-rotating black hole are well known, e.g. see \cite{Damour:Schafer:1988,Schafer:Wex:1993}.
Slow-motion or weak-field approximations (pN approximations) of highly analytical or involved calculations allows one to have 
 a meaningful interpretation of the results. In the references e.g. \cite{Damour:Schafer:1988}, \cite{Konigsdorffer:Gopakumar:2005},
 and \cite{Tessmer:Hartung:Schafer:2012} the motion of spinning binary systems were studied analytically and periastron
advances were calculated to higher pN orders. 
The role played by the Carter constant \cite{Carter:1968} in various problems of motion related with rotating black-holes is
an interesting subject, see \cite{Flanagan:Hinderer:2007,Will:2009}, where Carter-like constants are introduced. This procedure will also
be necessary for our calculations, for which we use Hill-inspired canonical variables  \cite{Hill:1913,Izsak:1963,Aksnes:1972} being better
suited for the problem in hand than the often applied Boyer-Lindquist coordinates \cite{Boyer:Lindquist:1967}.
 These canonical variables also appear in restricted form in a most recent article on binary-spin calculations
 for the special case of circular motion \cite{Tessmer:Steinhoff:Schafer:2013}.

We, for the first time, without restricting to any special case, calculate the orbital period, the intrinsic periastron
advance, and the precession of the orbital plane of a test-mass in an inclined, generic orbit in a Kerr spacetime field approximated
to second pN order for the monopole interaction, next-to-leading pN order linear in the spin parameter $a$, and the more important leading pN
order quadratic in $a^2$ between the spinless test-mass and its spinning massive companion. 
For a highly inclined orbit with high eccentricity one needs to carefully handle the influence of the Carter constant which is absent in the case of equatorial orbits.
We approach this problem by the most transparent way possible, which is, by starting from an invariant scalar action in canonical variables allowing a
 clear definition and understanding of the observables derived from it.
For our calculations we use SI units. 
Our perturbative method for making calculations feasible is the post-Newtonian approximation 
technique, which expands field equations and equations of motion (EOMs) in inverse powers of c, the speed of light.
Hence, if a term carries the power $\frac{1}{c^{n}}$, with $n\in\mathbb{N}_{0}$ it is referred to as being of the $\frac{n}{2}$th-pN order or $\frac{n}{2}$pN
 in short where 0pN is the 
usual Newtonian order.

\section{The underlying action principle}

The motion of the test-mass is governed by the binding energy, $E_{b}$, which is conserved, and defines a Hamiltonian for
generating the equations of motion. The total energy $E$ of the system is given by
\begin{align}
 E=m c^2+E_{b}
\end{align}
with $m$ being the rest-mass of the test-body. 
%Notice that for a true binary system we would have to take the reduced mass
%$\mu=\frac{mM}{m+M}$ instead of $m$ with M being the mass of the massive companion, from which the gravitational field is exerted,
%but because $m\ll M$ we get $\mu=m$. 
We calculate the binding energy to $O(c^{-5})$ by writing the Hamiltonian of the test-mass in the Kerr metric expanded to quadratic-in-spin
and then perturbatively solve for the energy, $E$ (expression shown below). Our specific calculation parallels the one
given in \cite{Fujita:Hikida:2009,Hackmann2:2010}. We also use the Hamilton-Jacobi equation
\begin{align}
 H\left(x^{\mu},\frac{\partial S}{\partial x^{\mu}}\right)+\frac{\partial S}{\partial\tau}=0
\end{align}
with $x^{\mu}\in\{t,r,\phi,\theta\}$ labelling the Boyer-Lindquist coordinates, $S$ being the underlying action,
 $\tau$ the proper time parameter and
\begin{align}\label{Hamcon}
H\equiv\frac{1}{2}g^{\mu\nu}p_{\mu}p_{\nu}=-\frac{1}{2}m^2c^4
\end{align}
 is the constrained
Hamiltonian of the system. The momenta are given by $p_{\mu}=\frac{\partial S}{\partial x^{\mu}}$ with the first two constants of motion,
 $p_{t}=-E$, the negative total energy, and $p_{\phi}=L_{z}$, the z-component of $m$'s angular momentum. Using a separation ansatz for the $r$- and $\theta$-dependent part
 of the action
$S^{r\theta}(r,\theta)=S^{r}(r)+S^{\theta}(\theta)$ the Hamilton-Jacobi equation can be further separated by introducing the third constant of motion, 
the Carter-like constant, $\mathcal{C}$,
 cf. \S 48 in \cite{Landau:Lifshitz:Vol1}, defined as
\begin{align}
 \mathcal{C}^2&:=L^2+\alpha^2a^2\cos^2\theta\qquad\text{with}\label{carter1}\\
\alpha^2&=m^2c^2\left(1-\frac{E^2}{m^2c^4}\right)\frac{r_{S}^2}{4}\\\nonumber
&=-\frac{2E_{b}}{mc^2}
\left(\frac{GMm}{c}\right)^2+\mathcal{O}(E_{b}^2)\,.
\end{align}
We only consider leading order terms in $a^2$, therefore, we can drop the $E_{b}^2$-terms being of higher pN order or next-to-leading-order. Notice that
$\alpha^2$ is positive because $E_{b}$ is negative for bound orbits. The remaining constrained equation following from (\ref{Hamcon})
 can be solved perturbatively for $E$ or $E_{b}$ respectively
yielding
\begin{align}
\begin{aligned}\label{Eb}
 &E_{b}=\frac{p_{r}^2}{2m}+\frac{\mathcal{C}^2}{2r^2m}-\frac{GMm}{r}+\frac{1}{c^2}\Bigg[G\left(-\frac{\mathcal{C}^2M}{2r^3m}-\frac{3Mp_{r}^2}{2rm}\right)\\
&-\frac{p_{r}^4}{8m^3}-\frac{\mathcal{C}^4}{8r^4m^3}-\frac{\mathcal{C}^2p_{r}^2}{4r^2m^3}-\frac{G^2M^2m}{2r^2}\Bigg]+\frac{1}{c^3}\frac{2aG^2L_{z}M^2}{r^3}\\
&+\frac{1}{c^4}\bigg[G\left(\frac{\mathcal{C}^4M}{8r^5m^3}+\frac{3\mathcal{C}^2Mp_{r}^2}{4r^3m^3}
+\frac{5Mp_{r}^4}{8rm^3}\right)-\frac{G^3M^3m}{2r^3}\\
&+G^2\left(-\frac{\mathcal{C}^2M^2}{4r^4m}+\frac{3M^2p_{r}^2}{4r^2m}\right)+\frac{\mathcal{C}^6}{16r^6m^5}+\frac{3\mathcal{C}^4p_{r}^2}{16r^4m^5}+\frac{p_{r}^6}{16m^5}\\
&+\frac{3\mathcal{C}^2p_{r}^4}{16r^2m^5}
+a^2G^2\left(-\frac{L_{z}^2M^2}{2r^4m}+\frac{M^2p_{r}^2}{2r^2m}\right)\bigg]+\mathcal{O}\left(\frac{1}{c^6}\right)
\end{aligned}  
\end{align}
with the dimensionless spin-$\mathcal{S}$ parameter $a=\mathcal{S}c/GM^2$ within in the range $a\in[-1,1]$, $r$ and $p_{r}$ being the radial
 coordinate and its canonical conjugate momentum respectively.
We use a coordinate system that has canonical Hill-inspired variables $(p_{r}, L, L_{z}; r, u, \Omega)$, with $u$ being
 the true anomaly, i.e.,
the angle between the position vector $\vct{r}=r\vct{n}$ and the direction of the
ascending node $\vct{N}=\left(\vct{e}_{z}\times\vct{L}\right)/|\vct{e}_{z}\times\vct{L}|$: $\cos u=\vct{n}\cdot\vct{N}$ and $\sin u=\vct{n}\cdot\vct{W}$ 
with $\vct{W}=\vct{L}\times\vct{N}/L$ (use of flat-space geometry is made). 
$\Omega$ is the angle of the ascending node as measured from the
$x$-axis of the non-rotating orthonormal basis ($\vct{e}_{x},\vct{e}_{y},\vct{e}_{z}$):
 $\cos\Omega=\vct{N}\cdot\vct{e}_{x}$ and $\sin\Omega=\vct{N}\cdot\vct{e}_{y}$, measuring the precession of the orbital
plane about $L_{z}$ which is tilted from the equatorial plane with the inclination angle $i$, the 
angle between the background $z$-axis and the orbital angular momentum vector $\mathbf{L}$, so $L_{z}=L\cos i$.
%
%We use a coordinate system that has canonical Hill-type variables $(p_{r}, L, L_{z}; r, u, \Omega)$, with $u$ being the angle swept by the test mass in the orbital plane, which is perpendicular to $\vct{L}$, and $\Omega$ is the angle that measures the precession of the orbital plane about $L_{z}$, satisfying 
The standard canonical Poisson bracket relations are
\begin{align}
 \{r,p_{r}\}=\{u,L\}=\{\Omega,L_{z}\}=1
\end{align}
with all other brackets being zero. In contrast to the classic Hill variables $(\dot r, G, H; r, \tilde{u}, \tilde{\Omega})$
 with $\vct{G}=\vct{r}\times\vct{v}$ ($\vct{v}$ being the coordinate velocity vector), 
$H=G_{z}$ and $(\tilde{u},\tilde{\Omega})$ being
conjugate to $\vct{G}$ and $H$ respectively, we work with the momentum and its derived quantities
$p_{r}=\vct{p}\cdot\vct{r}/r$, and $L=|\vct{L}|$ with $\vct{L}=\vct{r}\times\vct{p}$, i.e. our $(u,\Omega)$ differs from $(\tilde{u},\tilde{\Omega})$.
 They are equal in the
Newtonian case (per unit mass), see \cite{Gopakumar:Schafer:2011} for variable relations between these different frames.
Notice that our orbital plane is defined as being orthogonal to $\vct{L}$ and not to $\vct{G}$, where $\vct{G}$ is often denoted 
by $\vct{L}_N$ apart from a mass parameter.   
The corresponding action integral about a complete revolution from periastron to periastron reads, omitting $\tau$, 
\begin{align}\label{actionL}
\begin{aligned}
 S&=S(E_{b},L_{z},\mathcal{C}, P,\Phi, U)\\
 &=-E_{b} P+L_{z}\Phi+\oint p_{r}\dd r+\oint L\dd u\,.
\end{aligned}
\end{align}
It depends on the three constants of motion, $E_{b},L_{z},\mathcal{C}$, and three orbital-completed variables
\begin{align}
 P=\oint \dd t\,,\quad \Phi=\oint\dd\Omega\,,\quad\text{and}\quad U=\oint\dd u\,.
\end{align}
$P$ being the orbital period, $\Phi$ the precession of the orbital plane per revolution, and U the intrinsic periastron advance. 
The relation between the Hill-inspired variable $u$ and the Boyer-Lindquist coordinate $\theta$ is, cf. \cite{Damour:Schafer:1988,Schneider:Cui:2005},
\begin{align}
 \sin u\sin i = \cos\theta\, .
\end{align}
%where $i$ is the inclination angle between the background $z$-axis and the orbital angular momentum vector $\mathbf{L}$, i.e. , $\cos i=\frac{L_{z}}{L}$. 
 % new from SH but already present
%\textcolor{blue}{Solving Eq. (\ref{Eb}) for $p_r$ and Eq. (\ref{carter1}) for $L$ allows the straightforward calculation of the action, Eq. (\ref{actionL}). Within the context of our leading order calculation, $L$ in $\cos i$ can be replaced by $\mathcal{C}$.}
% 
Solving Eq. (\ref{Eb}) for $p_r$ and Eq. (\ref{carter1}) for $L$ allows the straightforward calculation of the action, Eq. (\ref{actionL}). 
Within the context of our leading order calculation, $L$ in $\cos i$ can be replaced by $\mathcal{C}$.
The next step is to calculate the remaining integrals in the action (\ref{actionL}). We start with
\begin{align}
\begin{aligned}
 &\oint L\dd u=\oint\sqrt{\mathcal{C}^2-\alpha^2a^2\sin^2i\,\sin^2u}\,\dd u\\
             &=\oint \mathcal{C}\dd u-\oint\frac{1}{2\mathcal{C}}a^2\alpha^2\sin^2 i\,\sin^2 u\,\dd u+\mathcal{O}(a^4)\\
            &=\mathcal{C}U-\frac{\pi}{2\mathcal{C}}a^2\alpha^2\left(1-\frac{L_{z}^2}{\mathcal{C}^2}\right)+\text{\text{NLO}-$a^2$-\text{terms}}\,.
\end{aligned}
\end{align}
Since the integrand is of leading-order in $a^2$ we integrate over a closed Newtonian orbit, i.e., we integrate from $0$ to $2\pi$
 and drop all next-to-leading-order (NLO) corrections. The action, adapted to our problem, reads
\begin{align}
 S=-E_{b} P+L_{z}\Phi+\oint p_{r}\dd r+\mathcal{C}U-\frac{\pi}{2\mathcal{C}}a^2\alpha^2\left(1-\frac{L_{z}^2}{\mathcal{C}^2}\right)\,.
\end{align}
The action principle tells us
\begin{align}
 \frac{\partial S}{\partial\mathcal{C}}=\frac{\partial S}{\partial L_{z}}=\frac{\partial S}{\partial E_{b}}=0
\end{align}
giving rise to formulas for the orbital elements,
\begin{align}
 U&=-\frac{\partial}{\partial\mathcal{C}}\oint p_{r}\dd r-\frac{a^2\pi\alpha^2(E_{b})}{2\mathcal{C}^2}
\left(1-\frac{3L_{z}^2}{\mathcal{C}^2}\right)\,,\\
 \Phi&=-\frac{\partial}{\partial L_{z}}\oint p_{r}\dd r-a^2\pi\alpha^2(E_{b})\frac{L_{z}}{\mathcal{C}^3}\,,\\
 P&=\frac{\partial}{\partial E_{b}}\oint p_{r}\dd r+\frac{a^2\pi G^2M^2m}{c^4\mathcal{C}}\left(1-\frac{L_{z}^2}{\mathcal{C}^2}\right)\,.
\end{align}
The integral over $p_{r}$ can be evaluated perturbatively in inverse powers of $c$,
 where $p_{r}$ is evaluated by inverting Eq. (\ref{Eb}) to $O(c^{-5})$.
To make the integration easy we choose the method of contour integration devised
by Sommerfeld in \cite{Sommerfeld:1951},
\begin{align}
 \oint p_{r}\dd r=-2\pi i\left[\text{Res}_{r=0}(p_{r})+\text{Res}_{r=\infty}(p_{r})\right]
\end{align}
The result reads, also see \cite{Damour:Schafer:1988,Damour:Jaranowski:Schafer:2000:2}:
\begin{align}
\begin{aligned}
 &\frac{1}{2\pi}\oint p_{r}\dd r=-\mathcal{C}+\frac{GMm}{\sqrt{-\frac{2E_{b}}{m}}}+\frac{1}{c^2}\bigg(\frac{3G^2M^2m^2}{\mathcal{C}}\\
&+
\frac{15GMm}{4}\sqrt{\frac{-E_{b}}{2m}}\bigg)-\frac{1}{c^3}\frac{2aG^3L_{z}M^3m^3}{\mathcal{C}^3}\\
&+\frac{1}{c^4}\Bigg[\frac{15E_{b}G^2M^2m^4}{2\mathcal{C}}+\frac{35G^4M^4m^4}{4\mathcal{C}^3}\\
&+\frac{35E_{b}GM}{32}\sqrt{-\frac{E_{b}}{2m}}
+a^2\bigg(E_{b}G^2\left(-\frac{M^2m}{2\mathcal{C}}+\frac{L_{z}^2M^2m}{2\mathcal{C}^3}\right)\\
&-\frac{G^4M^4m^4}{4\mathcal{C}^3}
+\frac{3G^4L_{z}^2M^4m^4}{4\mathcal{C}^5}\bigg)\Bigg]\\
&-\frac{a}{c^5}\left(\frac{12E_{b}G^3M^3L_{z}m^2}{\mathcal{C}^3}+\frac{21G^5M^5L_{z}m^5}{\mathcal{C}^5}\right)\,.
\end{aligned}
\end{align}
The two periastron shifts, $\Delta\bar{U}$ and $\Delta\bar{\Phi}$, the intrinsic and the one related with the precession 
of the orbital plane, respectively, are then giving by
\begin{align}\label{u}
\begin{aligned}
 \Delta \bar{U}&=\frac{1}{2\pi}(U-2\pi)=\frac{3G^2M^2m^2}{c^2\mathcal{C}^2}-\frac{6aG^3M^3m^3}{c^3\mathcal{C}^3}\cos i\\
               &+\frac{1}{c^4}\Bigg[\frac{15E_{b}G^2M^2m}{2\mathcal{C}^2}+\frac{105G^4M^4m^4}{4\mathcal{C}^4}\\
&+a^2\frac{3G^4M^4m^4}{4\mathcal{C}^4}\left(5\cos^2 i-1\right)\Bigg]\\
&-\frac{aG^3M^3m^2}{c^5\mathcal{C}^3}\left(36E_{b}+105\frac
{G^2M^2m^3}{\mathcal{C}^2}\right)\cos i
\end{aligned}
\end{align}
and
\begin{align}\label{phi}
\begin{aligned}
 \Delta\bar{\Phi}&=\frac{1}{2\pi}\Phi=\frac{2\bar{a}G^3M^3m^3}{c^3\mathcal{C}^3}+\frac{a^2}{c^4}\cos i
\left(-\frac{3G^4M^4m^4}{2\mathcal{C}^4}\right)\\
&\qquad+\frac{a}{c^5}\left(\frac{12E_{b}G^3M^3m^2}{\mathcal{C}^3}+\frac{21G^5M^5m^5}
{\mathcal{C}^5}\right)\,.
\end{aligned}
\end{align}
With Eqs. (\ref{u}) and (\ref{phi}) we reproduce a corresponding expression in \cite{Barker:OConnell:1979} in the test-mass limit, which results
 from the quadrupole deformation of the central
object, see Eq. (3.5) therein.
We also recover the usual pN result when the motion takes place in the equatorial plane.
In this case, $\mathcal{C}=L=L_{z}$ and our precession angle $\Phi$ agrees to leading pN order with Eq. (53) in \cite{Sibgatullin:2001},
 where it is called the nodal precession rate. Furthermore, the leading-order linear-in-$a$ terms of $\Delta\bar{U}$ and $\Delta\bar{\Phi}$ fully agree with
the results in \cite{Damour:Schafer:1988} for the case of an inclined orbital plane when referred to the test-mass limit.
For the period we have
\begin{align}\label{period}
\begin{aligned}
 P&=\frac{2\pi GM}{\left(-2E_{b}/m\right)^{3/2}}\bigg[1-\frac{1}{c^2}\frac{15}{4}\frac{E_{b}}{m}\\
&+
\frac{1}{c^4}\left(\frac{15}{2}\frac{GMm}{\mathcal{C}}\left(\frac{-2E_{b}}{m}\right)^{3/2}-\frac{105}{32}\frac{E_{b}^2}{m^2}\right)
\\
&-\frac{12aG^2M^2m^2}{c^5\mathcal{C}^2}\left(\frac{-2E_{b}}{m}\right)^{3/2}\cos i\bigg]\,.
\end{aligned}
\end{align}
In the case of equatorial motion the resulting shift, $\Delta\tilde\Phi$, (for this degenerate case) is given by
\begin{align}\label{phig}
\begin{aligned}
\Delta\tilde\Phi&=\Delta \bar{U}+\Delta\bar{\Phi}=-1-\frac{1}{2\pi}\left(\frac{\partial}{\partial \mathcal{C}}+\frac{\partial}{\partial L_{z}}\right)\oint p_{r}\dd r\\
                &=\frac{3G^2M^2m^2}{c^2\mathcal{C}^2}-\frac{4aG^3M^3m^3}{c^3\mathcal{C}^3}
               \\&+\frac{1}{c^4}\bigg[\frac{15E_{b}G^2M^2m}{2\mathcal{C}^2}+\frac{105G^4M^4m^4}{4\mathcal{C}^4}+a^2\frac{3G^4M^4m^4}{2\mathcal{C}^4}\bigg]\\
                &+\frac{a}{c^5}\left(-\frac{24E_{b}G^3M^3m^2}{\mathcal{C}^3}-\frac{84G^5M^5m^5}{\mathcal{C}^5}\right)\,.
\end{aligned}
\end{align}
This quantity has been calculated by other methods \cite{Tessmer:Hartung:Schafer:2012} whose result coincides with ours, also see \cite{Sibgatullin:2001} to
leading-pN-order. Notice that one can replace the Carter-like constant $\mathcal{C}$ by $L$ in
 the Eqs. (\ref{u})-(\ref{phig}) because the $a^2$-correction in $\mathcal{C}$
 is shifting the pN order of the corresponding terms beyond our considerations. 
The periastron shifts and the orbital period in Eqs. (\ref{u})-(\ref{phig}), are directly observable, and coordinate-invariant or gauge-invariant quantities.
 Their measurements impose three conditions on the experimental values of the gauge-invariant conserved quantities $E_b$, $\mathcal{C}$, and $L_z$,
 and constant parameters, $M$, $\mathcal{S}$, and $m$ provided the direction of the spin vector of the central object is known. A detailed treatment
 of the measurement of those directly observable quantities in binary pulsar systems has been performed by \cite{Damour:Schafer:1988} up to second post-Newtonian order. 
A convincing interpretation of the measured data needs to take into account also the propagation properties of the detected light 
emitted from the system in question. It is worthwhile pointing out that the notions of semi-major axis or eccentricity or semi-latus rectum
 of a relativistic orbit are not gauge invariant but depend on the chosen coordinate system beyond the Newtonian level. To leading order in Eqs. (\ref{u})-(\ref{phig}),
the involved physical quantities need to be known to Newtonian precision only.

\acknowledgments
S.H. thanks Eva Hackmann, Johannes Hartung, and Manuel Tessmer for useful discussions. 
This work was supported in part by the Deutsche Forschungsgemeinschaft (DFG) through
SFB/TR7 ``Gravitational Wave Astronomy'', GRK 1523 ``Quantum and Gravitational Fields'', and European Research Council Starting Grant No. 202996.

%\ifnotprd
\bibliographystyle{utphys}
%\fi

%\ifarxiv
 %run: cp LieTrafosAndCanonicAngles.bbl LieTrafosAndCanonicAngles_refs_arxiv.tex

\providecommand{\href}[2]{#2}\begingroup\raggedright\endgroup

%\bibliography{../references}
%\fi

\end{document}